# Creation of Magnetic Skyrmion Bubble Lattices by Ultrafast Laser in Ultrathin Films

Soong-Geun Je[1,2,*,†], Pierre Vallobra[2], Titiksha Srivastava[1], Juan-Carlos Rojas-Sánchez[2], Thai Ha Pham[2], Michel Hehn[2], Gregory Malinowski[2], Claire Baraduc[1], Stéphane Auffret[1], Gilles Gaudin[1], Stéphane Mangin[2], Hélène Béa[1], Olivier Boulle[1,†]

[1]Université Grenoble Alpes, CNRS, CEA, Grenoble INP, INAC-SPINTEC, 38000 Grenoble, France

[2]Institut Jean Lamour, CNRS UMR 7198, Université de Lorraine, Nancy F-54500, France

[*]Current address: Center for X-ray Optics, Lawrence Berkeley National Laboratory, Berkeley, California 94720, USA; Department of Emerging Materials Science, Daegu Gyeongbuk Institute of Science & Technology, Daegu 42988, Republic of Korea

[†]email: soonggeun.je@gmail.com; olivier.boulle@cea.fr

## Abstract

Magnetic skyrmions are topologically nontrivial spin textures which hold great promise as stable information carriers in spintronic devices at the nanoscale. One of the major challenges for developing novel skyrmion-based memory and logic devices is fast and controlled creation of magnetic skyrmions at ambient conditions. Here we demonstrate controlled generation of skyrmion bubbles and skyrmion bubble lattices from a ferromagnetic state in sputtered ultrathin magnetic films at room temperature by a single ultrafast (35-fs) laser pulse. The skyrmion bubble density increases with the laser fluence, and it finally becomes saturated, forming disordered hexagonal lattices. Moreover, we present that the skyrmion bubble lattice configuration leads to enhanced topological stability as compared to isolated skyrmions, suggesting its promising use in data storage. Our findings shed light on the optical approach to the skyrmion bubble lattice in commonly accessible materials, paving the road toward the emerging skyrmion-based memory and synaptic devices.

Ultrathin magnetic films have been the subject of intense research for data storage applications such as domain-wall (DW) racetrack memory [1] and magnetic random access memory [2,3]. Recently, considerable attention has been given to ultrathin multilayers composed of heavy metal (HM) and/or oxide layers in contact with ultrathin *3d* transition metals (TM). In this system with perpendicular magnetic anisotropy (PMA), the strong spin-orbit coupling and the structural inversion asymmetry are found to lead to unexpected rich physics such as the spin-orbit torque (SOT) [4-7] and the interfacial Dzyaloshinskii-Moriya interaction (DMI) [8-14]. These phenomena immediately became key ingredients for the ultrathin film-based spintronic devices, including efficient current-induced manipulation of magnetization as well as fascinating nontrivial and noncollinear magnetization structures [9-19]. One of the most prominent examples, promoted by both the SOT and DMI, is magnetic skyrmions in ultrathin magnetic materials [20-23] that possess rich physical and topological properties [24,25] and prospects of applications [26,27].

Skyrmions were first observed in bulk B20 chiral magnets at low temperature [28-31] where Bloch-like skyrmions are stabilized by the DMI [32,33] due to the non-centrosymmetric crystal structure. Later, they were found in epitaxial ultrathin films at low temperature [34-36] and more recently at room temperature in sputtered HM/TM ultrathin films [20-23,38-40]. In these ultrathin films, the interfacial DMI [37], which arises from the asymmetric interfaces, leads to Néel skyrmions with fixed chirality [20]. It was recently shown that these chiral Néel skyrmions can efficiently be driven by electric currents using the SOT [21,23,41-46]. This has suggested novel concepts of memory devices that would combine very high-density data storage, fast access time and low-power consumption by exploiting topologically stable nanometre-scale skyrmions as information carriers [27,47]. Moreover, skyrmion-based neuromorphic and stochastic computing schemes [48,49] have lately been proposed, expanding their use beyond data storage applications.

For such devices, one essential requirement is to achieve low power, fast and controllable writing of skyrmions. To date, various room-temperature skyrmion creation schemes have been reported. With well-tuned material parameters, dc magnetic fields allow the system to reach the skyrmion state by shrinking pre-existing small worm domains [38,41], but the contracted skyrmions are generally placed at the strong pinning sites [50]. Creation schemes using electric currents, accompanied by the SOT [23,51,52], thermal assistance [43,46] and Oersted field [21], have been proposed. In the electric current-induced scheme, however, power consumption is generally high, and special geometric setups are required [23,51]. Moreover, the nucleation currents are generally higher than the driving currents [46,51], resulting in crosstalk between writing and driving operations. Electric field-induced manipulation of skyrmions has been demonstrated [40], but the operation time can be limited by the RC time constant. Besides, achieving both the sensitive response to the electric field and high thermal stability of skyrmions imposes significant constraints on the choice of materials.

Ultrafast all-optical manipulation of magnetization, observed for a wide range of materials [53-55], can provide a breakthrough in the fast and efficient writing of skyrmions. So far, it has been shown that the direction of the magnetization can be set by laser in a helicity-dependent or helicity-independent way [56,57]. Recently, it was also reported that the ultrafast laser pulses drive domain walls in cooperation with temperature gradient [58]. All those achievements might point towards the possibility of optical manipulation of skyrmions. Moreover, as laser acts locally, this could be particularly relevant to the skyrmion synapse device [48] where creating clusters of skyrmions in a specific region of the device is a key for high resolution synaptic weights.

Recently, it was theoretically proposed that local heating using laser excitation would allow the nucleation of magnetic skyrmions and antiskyrmions on a sub-ns time scale [59]. Experimentally, several works demonstrated the laser-induced nucleation of non-trivial magnetic textures. However, the experiments were carried out in materials exhibiting non-homochiral magnetic textures [60] or in a bulk material at low temperature [61], which are not relevant for devices based on skyrmion manipulation. Thus, the demonstration of the optical generation of homochiral magnetic skyrmions at room temperature would be an important milestone towards the manipulation of magnetic skyrmion in memory and logic devices and would open a path for an ultrafast, low energy consuming writing scheme.

Here we report on the ultrafast laser-induced generation of skyrmion bubbles and skyrmion bubble lattices in an ultrathin ferromagnetic layer. An ensemble of skyrmion bubbles is created from a ferromagnetic state by a 35-fs single laser pulse with low fluences. With the increasing laser fluence, the transition from an isolated skyrmion bubble state to a skyrmion bubble lattice state with controllable bubble density is observed. The laser heating-induced skyrmion bubble lattice formation is explained in the framework of the analytic bubble and bubble lattice models and micromagnetic simulations. Our results highlight the role of thermal effect in the creation of the skyrmion bubble lattice. It also offers an alternative way of ultrafast excitation and massive writing of clusters of skyrmions that decouples the writing operation from the skyrmion driving operation in the skyrmion-based shift register and that could be used in the emerging skyrmion-based synaptic devices.

For this study, a Ta(30 Å)/Fe$_{72}$Co$_8$B$_{20}$(wedge, 7.5 Å to 10.3 Å)/TaO$_x$(12 Å, naturally oxidized) trilayer film is grown on a glass substrate. This stack structure is chosen because it has the interfacial DMI large enough to stabilize chiral bubbles, thus enabling the SOT-induced skyrmion bubble motion [23,42,45,62]. More detailed study on the DMI in the present structure can be found in Ref. 62. Magnetic domains are imaged using magneto-optical Faraday effect, allowing observation during laser illumination (Fig. 1a). Fig. 1b-e show the magnetic hysteresis loop and corresponding domain images taken at the FeCoB thickness of about 9 Å. Bright and dark contrast correspond to magnetization pointing up (+z) and down (-z), respectively. At zero magnetic field, labyrinthine domains are observed (Fig. 1d). The spontaneous formation of periodic up and down domains is explained by the long-range dipolar field that also favours the formation of skyrmions and bubbles. We note, however, that the labyrinthine domains never transform into skyrmion bubbles by applying external magnetic fields in the films. The stripe domains pointing antiparallel to the external field just get narrow (Fig. 1e) and the film is finally saturated (Fig. 1c). This behaviour is observed over the whole FeCoB thickness range (7.5-10.3 Å), and thus we will mainly discuss the data obtained for 9 Å thick film unless otherwise mentioned.

The labyrinthine domain pattern allows us to determine the DMI strength in the present film. In the parallel stripe domain model [63], the labyrinthine domain width $L_0$ is given by $L_0 = d\frac{\pi}{2}\exp\left(\frac{\sigma_{DW}}{4\lambda d} - \frac{1}{2}\right)$ as a function of the film thickness $d$, the domain wall energy $\sigma_{DW}$ and the dipolar energy constant $\lambda = \frac{\mu_0}{4\pi}M_S^2$ with the saturation magnetization $M_S$. In Fig. 1d, we determine $L_0 = 1.85$ μm using a fast Fourier transform. The experimental $d(= 9$ Å$)$ and $M_S(= 1$ MA/m$)$ lead to $\sigma_{DW} = 2.76$ mJ/m$^2$. Note that this value is smaller than $4\sqrt{A_{ex}K_{eff}} = 3.25$ mJ/m$^2$, calculated from the experimental effective anisotropy $K_{eff}$ and the typical exchange stiffness $A_{ex} = 12$ pJ/m [62], implying the DMI contribution $(-\pi|D|)$ to $\sigma_{DW}(= 4\sqrt{A_{ex}K_{eff}} - \pi|D|)$. We estimate the DMI energy $D = 0.16$ mJ/m$^2$, in agreement with the reported value [62]. The stripe width with respect to the external magnetic field also leads to the same $D$ value, and a micromagnetic simulation predicts the DMI-stabilized chiral Néel domain wall in the present film (supplementary information).

**Ultrafast laser-induced creation of skyrmion bubbles**

The influence of the laser pulse on the magnetic system is then investigated. First, we prepare a saturated initial state (marked by c in Fig. 1b) by saturating the film and reducing the magnetic field down to +0.31 mT to suppress natural nucleation of reversed domains. The initial state is then exposed to a linearly polarized 35-fs single laser pulse with a spatial full-width half-maximum (FWHM) of ~60 μm. Fig. 2a shows the magnetic configuration after the laser excitation. Interestingly, bubble domains of ~2.5 μm in diameter with the magnetization antiparallel to the magnetic field are created. Bubbles with opposite magnetization are also generated for the opposite initial state and field as shown in Fig. 2b. It is worth noting that the bubble nucleation is achieved by excitation about a thousand times shorter than the recent demonstration using an electric current [51]. As mentioned above, since the domain walls in our film are homochiral Néel walls due to the presence of DMI, the created bubbles are chiral skyrmion bubbles that bear the topological similarity to skyrmions.

To see whether there is laser helicity dependence, such as the all-optical helicity dependent switching [53,54], the bubble generation is also tested for circularly polarized lights (right polarized $\sigma^+$ and left polarized $\sigma^-$) as shown in Fig. 2c-d. The two circularly polarized lights result in no noticeable difference, implying that the laser-induced skyrmion bubble generation is mainly attributed to laser-induced transient heating. The bubble lattice area is 2 times smaller than the FWHM of the laser. We attribute this to a Gaussian intensity profile of the beam, and this indicates that there is a fluence threshold above which skyrmion bubbles are nucleated. As there is no helicity dependence, we will focus on the case of the linearly polarized light for simplicity.

For a thicker FeCoB (10 Å) region, narrower labyrinthine domains are observed (Fig. 2e). Accordingly, smaller chiral bubbles (<1 μm in diameter) are nucleated by laser (Fig. 2f). This might indicate that the laser heating-induced method can be applicable to the generation of much smaller skyrmions that are beyond the resolution of the current experimental setup.

Moreover, this method can also be used for single skyrmion creation, provided that the laser is much more focused with a micro-sized beam spot [59] or the laser is a vortex beam [64]. Hereafter, we focus on the sample region where the larger bubbles are generated for the convenience of image analysis.

**Skyrmion bubble lattice created by laser heating**

Next, we examine the dependence of the laser fluence on the skyrmion bubble generation. Fig. 3a-e present the ensemble of bubbles generated by the laser for various fluences in the presence of $\mu_0 H_z$ = +0.28 mT. To obtain bubbles on a large area, laser pulses with a repetition rate of 5 kHz were swept on the film ±20 μm in the *x* and *y* directions (dashed circle and arrows in Fig. 3a). At low laser fluences (Fig. 3a-b), individual bubbles are randomly nucleated, resulting in scattered bubbles. As the laser fluence increases, however, the bubble distribution becomes denser and starts to completely cover all the area swept by the laser. The bubbles then finally form a weakly ordered lattice structure. A distinct feature can be found at higher laser fluences (Fig. 3c-e). In those cases, the size of the bubbles in the central region, where bubbles are surrounded by bubbles, is smaller than that of bubbles lying on the boundary of the bubble area.

Fig. 3f shows the bubble density, the number of bubbles in a unit area, as a function of the laser fluence. With increasing laser fluence, the bubble density increases, and it is finally saturated at higher laser fluence. Note that when exposed to laser pulses of much higher fluences than those required for the saturation shown in Fig. 3f, the film properties irreversibly changed, implying the link between the onset of densely packed bubbles and the significant heating effect.

The above observations can be understood in the framework of the long-range dipolar repulsion between bubbles [65,66]. The increase in the laser fluence leads to the nucleation of an increased number of bubbles on a larger area. As a result, a larger resultant repulsive force is felt by the bubbles in the centre of the lattice, leading to larger bubble density and smaller bubble radii. For much more bubbles in the lattice, a saturation occurs as the outermost bubbles have a less and less effect on the bubble in the centre.

The bubble lattice is analysed using Delaunay triangularization [38] as exemplarily shown in Fig. 3g. The statistics of the nearest neighbour number ($N_{NN}$) and the angle of Delaunay triangles ($\theta_{DT}$) are plotted in Fig. 3h-i, respectively. At low laser fluences, the distributions are broad without distinct peaks. For higher laser fluences, however, the distributions show clear peaks at $N_{NN}$ = 6 and $\theta_{DT}$ = 60°, indicating that the laser-generated bubbles form a disordered hexagonal bubble lattice. The mean distance between nearest neighbours shows a decreasing trend with increasing laser fluence (Fig. 3j) that is in line with the dense bubble distribution in the central region.

**Analytic approach to the stabilization of the skyrmion bubble lattice**

First, we analyse the laser heating-induced skyrmion bubble generation in the framework of the analytic bubble energy model [40,63,67,68], which describes the stable bubble domain size and the energy barrier for its occurrence. Slightly different forms of equations in the various works stem from different treatments of magnetostatic energy. However, we note that they yield the same numerical results in our case (bubble radius ≫ film thickness). The energy of an isolated magnetic bubble of radius $R$ relative to the saturated state is written as $E_B = 2\pi R \lambda d \left[ \frac{\sigma_{DW}}{\lambda} - 4d \ln\left(\frac{8R}{d\sqrt{e}}\right) + Rh \right]$ where $h$ is the reduced external magnetic field $4\pi H_z/M_S$ and $e$ is Euler's number [63].

Using the previously determined $\sigma_{DW}$ and the experimental values, the bubble energies as a function of $R$ for various magnetic fields $h$ are plotted in Fig. 4a. For small $h$ (< 4.15×10$^{-9}$, $\mu_0 H_z$ < 0.33 mT), $E_B$ has two local energy minima at $R = 0$ and at some $R^*$ where the energy is lower than zero (i.e., global minimum). As discussed by Schott *et al*. [40], to reach this global minimum at $R^*$, a thermally nucleated nanoscale skyrmion around $R = 0$ needs to grow by overcoming the energy barrier ($E_n$, the local maximum). If $k_B T$ is comparable to the energy barrier, the bubble state can be achieved by simply applying an external field in materials with well-tuned parameters. On the other hand, if $E_n$ is large compared to $k_B T$, the spontaneous transition to the skyrmion bubble state cannot happen. In our case, however, laser heating provides sufficient energy to the system, enhancing the nucleation of the nanoscale skyrmion and, at the same time, enabling them to easily hurdle the energy barrier $E_n$. In Fig. 4a, we estimate $E_n$ to be very high, around 4×10$^{-21}$ J. For the typical laser fluence of 1.4 mJ/cm², the estimate of the energy on a nanoscale spin structure is of the order of 10$^{-17}$ J. This energy is large compared to the nucleation barrier energy and thereby allows bubbles to be reliably nucleated.

The skyrmion bubble lattice nucleation can then be understood as follows: as shown in Supplementary Movie 1, during laser pulse, heating almost demagnetizes the saturated area, creating fluctuating tiny domains. Immediately after the laser pulse, bubble domains that overcame the energy barrier $E_n$ stabilize with a radius around $R^*$. Due to the dipolar interaction, these bubbles repel each other and finally settle down in the quasi-hexagonal lattice formation. The energy of the hexagonal bubble lattice per unit volume is given by [63] $e_{BL} = \frac{4\pi R \lambda}{\sqrt{3} L_B^2} \left[ \frac{\sigma_{DW}}{\lambda} - 4d \ln\left(\frac{8R}{d\sqrt{e}}\right) + Rh + 6\pi d \sum_{k=0}^{\infty} S_k \left(\frac{R}{L_B}\right)^{3+2k} \right]$ where the last term accounts for the dipolar energy from neighboring bubbles with the coefficient $S_k$, which diminishes with increasing $k$, and $L_B$ is the distance between neighbouring bubble centres. By assuming that we have sufficient number of bubbles to build a hexagonal lattice, the minimum of $e_{BL}$ and the corresponding $R^*$ with respect to $L_B$ are plotted in Fig. 4b. There exists a global minimum point at $L_B^*$, implying that having the bubble spacing below $L_B^*$ is not preferred. Since $L_B^2$ is inversely proportional to the bubble density, the existence of $L_B^*$ explains the saturation of the bubble density in Fig. 3f. Moreover, the smaller $R^*$ (~1.1 μm) than that of the single bubble case ($R^*$ = 1.4 μm in Fig. 4a, $h$ = 3.52×10$^{-9}$) agrees with the observation that the size of the bubbles become smaller in the central area with increasing the number of bubbles.

**Micromagnetic simulations**

To better understand the stabilization of the skyrmion bubble lattice, we carried out micromagnetic simulations [69,70]. Although atomistic spin dynamics or the three-temperature model would be needed to describe the laser-induced demagnetization and the dynamics just after the laser pulse on the ps timescale [71,72], micromagnetic simulation allows one to capture most of the physics of the domain formation after the laser-induced demagnetization on a longer timescale. Besides, it should be noted that, since the micromagnetic simulations on the experimental length scales require an enormous computational time, we performed the simulation with magnetic parameters yielding smaller skyrmions in a small simulation area (2 × 2 μm$^2$) and a larger damping constant than the experimental value in a similar magnetic system [73], thus providing qualitative understanding. The parameter set for smaller skyrmions (see Methods) produces the labyrinthine domain pattern at remanence (Fig. 4c). Fig. 4d-j show the time evolution of the domain morphology. To mimic the laser heating area, the initial state (Fig. 4d) was prepared by relaxing the uniform magnetization state in the presence of randomly fluctuating fields in the central area (see Methods). The time evolution of individual energy terms is also investigated as shown in Fig. 4k.

The energy terms (Fig. 4k) with respect to simulation time $t$ exhibit remarkable features. First, the total energy very quickly decreases in a very short timescale ($t < 0.04$ ns, red triangle in Fig. 4k). The decrease is mainly due to the rapid decrease in the Heisenberg exchange energy, resulting in small up and down magnetization regions connected by in-plane magnetized regions. The presence of in-plane magnetized regions (encoded by white in Fig. 4d-e), which can also be deduced from the high anisotropy energy level (Fig. 4k), might imply the co-planar and non-collinear in-plane spin configuration that is crucial for the topological charge transition [74]. On this timescale, the chirality does not clearly appear in the domain morphology. However, soon after, domain walls possess the chiral nature around 0.2 ns (Fig. 4f), at which the DMI energy also shows a dip. After the rapid drop in the exchange energies, the uniaxial anisotropy plays a role and creates well-defined perpendicular domains (0.04 ns $< t <$ 5 ns, Fig. 4g and purple triangle in Fig. 4k), and consequently the demagnetization energy increases due to the reduced amount of in-plane magnetization components. After the domain morphology is clearly established in the central area, the demagnetization energy dominantly reduces the total energy by the repulsion [66] between skyrmions (Fig. 4h-i) and they are finally arranged in a hexagonal skyrmion lattice (Fig. 4j).

**Bubble-bubble repulsion**

In real materials, however, the formation of a hexagonal bubble lattice can be impeded by the presence of microstructural disorders, thereby leading to the disordered hexagonal lattice. To examine how the repulsive interaction works in real materials, we have measured the evolution of skyrmion bubble positions perturbed by laser in a disordered lattice. Fig. 5a-h show the sequence of images after every single laser pulse. The laser fluence is adjusted to minimize the bubble creation and thus allows us to track the single bubble behaviours. Bubbles are sometimes pushed out of the central area (red, cyan and pink circles) and are divided into two which expel each other while keeping apart from neighbouring bubbles (yellow circles). The motions after each laser pulse can be explained by the depinning and the dipolar field: bubbles are initially located at pinning sites which might act as local energy minima at the same time. Once a bubble is depinned by laser heating, then the repulsive dipolar forces from other bubbles push out the depinned bubble. For much higher laser fluence, more enhanced cooperation between the nucleation process and the repulsive process can drive the bubbles to more easily form the disordered hexagonal lattice structure (the process is more obviously shown in Supplementary Movie 2).

**Topological character of the skyrmion bubble lattice**

As the skyrmion bubbles in a lattice closely interact with each other, their response to an external magnetic perturbation can be different from that of a single bubble. To verify this, we first prepared a bubble lattice by sweeping the laser (Fig. 6a), and the magnetic field is then turned off (Fig. 6b). Interestingly, the initial closely packed bubbles hold their circular morphology and the lattice pattern while new stripe domains grow outside the lattice and cover the external area (Fig. 6b). When a reversed field is applied, the bubbles expand, but their expansion is limited by the adjacent bubbles, constructing a Voronoi-like network (Fig. 6c) which originates from the topological prevention of annihilation of paired Néel domain walls [75]. On the other hand, initially less packed bubbles (Fig. 6d) elongate or transform into stripe domains at zero field (Fig. 6e). When the field is reversed, the number of isolated domains, which carry the topology of their initial bubble states, decreases (Fig. 6f). This fact implies that the bubble lattice protects bubbles against perturbative magnetic fields that could be advantageous for preventing information loss in the outlook of memory and logic devices.

We systematically study how quickly the number of skyrmion bubbles decays with the external magnetic field perturbation depending on the initial number of bubbles. The magnetic field sequence is schematically shown in Fig. 6g. We first create the initial bubbles at the base field $\mu_0 H_{ini}$ of +0.28 mT. The initial number of bubbles $N_{ini}$ is controlled by tuning the laser fluence. The number of the bubbles $N_{pert}$ is counted every time when the field goes back to $\mu_0 H_{ini}$ after each magnetic field perturbation $\mu_0 H_{pert}$. As an indicator of the preservation of the overall topology, we introduce the retention ratio $RR = \frac{N_{pert}}{N_{ini}}$ and $RR$ is plotted with respect to $\mu_0 H_{pert}$ as shown in Fig. 6h.

This plot shows two distinct effects. First, in the positive $\mu_0 H_{pert}$ region where the magnetic field tends to shrink the bubbles, $RR$ exhibits no clear $N_{ini}$ dependence. In this case, the bubble-bubble interaction is suppressed so that the bubbles behave like a single bubble even in a lattice. On the other hand, in the negative $\mu_0 H_{pert}$ region where the magnetic field tends to expand the bubbles, $RR$ exhibits clear $N_{ini}$ dependence. The more bubbles we have, thus having the more packed lattice formation, the more the bubbles survive under the stronger $\mu_0 H_{pert}$, implying the role of the bubble-bubble interaction and thus enhanced topological stability. The faster decrease in $RR$ for the shrinking bubble case may indicate the lack of topological protection of isolated skyrmions due to the discrete nature of materials [76,77]. However, the twice larger $\mu_0 H_{pert}$ of the expanding case than that of the shrinking case (positive $\mu_0 H_{pert}$) means that the annihilation of the paired homochiral domain walls requires much higher energy (higher stability) than the skyrmion bubble annihilation. These results highlight the distinct topological feature of the closely packed skyrmion bubble lattice in terms of the immunity to the external interference and might provide important insight into storing information in the form of skyrmion lattices of different densities (Fig. 6i).

Finally, our results open intriguing prospects for the manipulation of skyrmions in skyrmion-based neuromorphic computing. Huang et al. [48] recently proposed the synaptic device where the synaptic weight corresponds to the number of skyrmions in a box. Our results show that potentiation and depreciation of the synapse, i.e., increase and decrease of the synaptic weight, can be achieved using respectively ultrafast laser pulses and magnetic fields. Indeed, the potentiation phase can be realized by the increasing skyrmion density (synaptic weight) with the laser fluence (Fig. 3f). On the other hand, the depreciation phase can be accomplished using magnetic field pulses as shown in Fig. 6h.

In this work, we demonstrated the ultrafast laser heating-induced generation of disordered hexagonal skyrmion bubble lattices in ultrathin magnetic films at room temperature. The skyrmion bubble density was controlled by laser fluence, suggesting the reliable means of manipulating skyrmion lattices. As this creation scheme is straightforward and free from complicated setups, it can be an efficient platform to quickly and widely search for the materials giving skyrmions at the film level. It can also be utilized for the writing operation in the skyrmion-based shift register (Fig. 6i) without disturbing the current-induced driving operations. Moreover, the control of the skyrmion density using laser pulses opens a promising outlook for skyrmion-based synaptic devices. While the manipulation of the single skyrmions has been the primary focus of interest, we presented enhanced stability of skyrmion bubble lattice, which might pave a new way toward the use of the ensemble of skyrmions for memory or logic devices.

**Methods**

**Sample preparation and experimental setup**

The Ta(30 Å)/Fe$_{72}$Co$_8$B$_{20}$(wedge)/TaO$_x$(12 Å, naturally oxidized) film was grown by DC magnetron sputtering. The wedged Fe$_{72}$Co$_8$B$_{20}$ thickness (nominal) is ranging from 7.5 Å to 10.3 Å. The wedged layer is made using off-axis deposition. The film is then annealed at 250 °C for 30 min to enhance PMA. For the laser excitation, we use a Ti: sapphire fs-laser (35-fs duration, 5 kHz repetition, 800 nm wavelength, a spatial FWHM of ~60 μm). A quarter-wave plate is used to change the helicity of the laser. The magnetic domains are imaged using a magneto-optic Faraday microscope.

**Micromagnetic simulations**

Micromagnetic simulations were performed using Object Oriented Micromagnetic Framework (OOMMF) with DMI package [69,70] on 2 × 2 m$^2$ area with a cell size of 5 nm. To produce skyrmions in the micro-sized simulation area, the following parameters are used for qualitative understanding: $D$ = 1.0 mJ/m$^2$, a uniaxial anisotropy $K$ = 6.8 × 10$^5$ J/m$^3$, $M_S$ = 1 MA/m, $A_{ex}$ = 12 pJ/m, $d$ = 0.9 nm, a damping constant $\alpha$ = 0.5 and a dc magnetic field of +5 mT. To mimic the thermally demagnetized region by laser heating, a saturated state is sufficiently relaxed in the presence of randomly fluctuating fields and the dc field with the same simulation parameters except for $K$ (2.3 × 10$^5$ J/m$^3$) and $A_{ex}$ (4 pJ/m). Then the relaxed magnetization area (1 × 1 μm$^2$) is embedded in the centre of the 2 × 2 μm$^2$ simulation area for the initial magnetization.

**Acknowledgements**

We appreciate helpful discussions with Dr Liliana D. Buda-Prejbeanu. This work was supported by the ANR-15-CE24-0009 UMAMI and by the ANR-Labcom Project LSTNM, by the Institut Carnot ICEEL for the project Optic-switch and Matelas and by the French PIA project Lorraine Université d'Excellence, reference ANR-15-IDEX-04-LUE. Experiments were performed using equipment from the TUBE-Daum funded by FEDER (EU), ANR, Région Grand Est and Metropole Grand Nancy. S.G.J thanks Dr Kyoung-Woong Moon for useful discussions. S.G.J. also acknowledges the support from the Future Materials Discovery Program through the NRF funded by the MSIP (2015M3D1A1070465) and from the DGIST R&D program (18-BT-02).

**References**

1. Parkin, S. S. P., Hayashi, M., Thomas, L. Magnetic domain-wall racetrack memory. Science 320, 190-194 (2008).

2. Ikeda, S. et al. Magnetic tunnel junctions for spintronic memories and beyond. IEEE Trans. Electron Devices 54, 991–1002 (2007).

3. Mangin, S. et al. Current-induced magnetization reversal in nanopillars with perpendicular anisotropy. Nature Mater. 5, 210–215 (2006).

4. Miron, I. M. et al. Perpendicular switching of a single ferromagnetic layer induced by in-plane current injection. Nature 476, 189–193 (2011).

5. Garello, K. et al. Symmetry and magnitude of spin-orbit torques in ferromagnetic heterostructures. Nature Nanotech. 8, 587–593 (2013).

6. Liu, L., Lee, O. J., Gudmundsen, T. J., Ralph, D. C. & Buhrman, R. A. Current-Induced Switching of Perpendicularly Magnetized Magnetic Layers Using Spin Torque from the Spin Hall Effect. Phys. Rev. Lett. 109, 096602 (2012).

7. Haazen, P. P. J. et al. Domain wall depinning governed by the spin Hall effect. Nature Mater. 12, 299-303 (2013).

8. Bode, M. et al. Chiral magnetic order at surfaces driven by inversion asymmetry. Nature 447, 190–193 (2007).

9. Thiaville, A., Rohart, S., Jué, É., Cros, V. & Fert, A. Dynamics of Dzyaloshinskii domain walls in ultrathin magnetic films. Europhys. Lett. 100, 57002 (2012).

10. Chen, G. et al. Novel Chiral Magnetic DomainWall Structure in Fe/Ni/Cu(001) Films. Phys. Rev. Lett. 110, 177204 (2013).

11. Je, S.-G. et al. Asymmetric magnetic domain-wall motion by the Dzyaloshinskii-Moriya interaction. Phys. Rev. B 88, 214401 (2013).

12. Pizzini, S. et al. Chirality-induced asymmetric magnetic nucleation in Pt/Co/AlO$_x$ ultrathin microstructures. Phys. Rev. Lett. 113, 047203 (2014).

13. Tetienne, J.-P. et al. The nature of domain walls in ultrathin ferromagnets revealed by scanning nanomagnetometry. Nature Commun. 6, 6733 (2015).

14. Emori, S., Bauer, U., Ahn, S.-M., Martinez, E. & Beach, G. S. D. Current-driven dynamics of chiral ferromagnetic domain walls. Nature Mater. 12, 611-616 (2013).

15. Ryu, K.-Su., Thomas, L., Yang, S.-H. & Parkin, S. Chiral spin torque at magnetic domain walls. Nature Nanotech. 8, 527-533 (2013).

16. Yang. S.-H., Ryu, K.-S. & Parkin, S. Domain-wall velocities of up to 750 m s−1 driven

by exchange-coupling torque in synthetic antiferromagnets. Nature Nanotech. 10, 221–226 (2015)

17. Garello, K. et al. Ultrafast magnetization switching by spin-orbit torques. Appl. Phys. Lett. 105, 212402 (2014).

18. Yoshimura, Y. et al. Soliton-like magnetic domain wall motion induced by the interfacial Dzyaloshinskii–Moriya interaction. Nature Phys. 12, 157–161 (2016).

19. Bogdanov, A. & Hubert, A. The stability of vortex-like structures in uniaxial ferromagnets. J. Magn. Magn. Mater. 195, 182–192 (1999).

20. Boulle, O. et al. Room-temperature chiral magnetic skyrmions in ultrathin magnetic nanostructures. Nature Nanotech. 11, 449–454 (2016).

21. Woo, S. et al. Observation of room temperature magnetic skyrmions and their current-driven dynamics in ultrathin Co films. Nature Mater. 15, 501–506 (2016).

22. Moreau-Luchaire, C. et al. Additive interfacial chiral interaction in multilayers for stabilization of small individual skyrmions at room temperature. Nat. Nanotech. 11, 444–448 (2016).

23. Jiang, W. et al. Blowing magnetic skyrmion bubbles. Science 349, 283 (2015).

24. Braun, H.-B. Topological effects in nanomagnetism: from superparamagnetism to chiral quantum solitons. Adv. Phys. 61, 1–116 (2012).

25. Nagaosa, N. & Tokura, Y. Topological properties and dynamics of magnetic skyrmions. Nature Nanotech. 8, 899–911 (2013).

26. Sampaio, J. et al. Nucleation, stability and current-induced motion of isolated magnetic skyrmions in nanostructures. Nature Nanotech. 8, 839–844 (2013).

27. Fert, A., Cros, V. & Sampaio, J. Skyrmions on the track. Nature Nanotech. 8, 152–156 (2013).

28. Mühlbauer, S. et al. Skyrmion lattice in a chiral magnet. Science 323, 915–919 (2009).

29. Yu, X. Z. et al. Real-space observation of a two-dimensional skyrmion crystal. Nature 465, 901–904 (2010).

30. Yu, X. Z. et al. Near room-temperature formation of a skyrmion crystal in thin-films of the helimagnet FeGe. Nature Mater. 10, 106–109 (2011).

31. Shibata, K. et al. Towards control of the size and helicity of skyrmions in helimagnetic alloys by spin-orbit coupling. Nature Nanotech. 8, 723–728 (2013).

32. Moriya, T. Anisotropic superexchange interaction and weak Ferromagnetism. Phys. Rev. 120, 91–98 (1960).

33. Dzyaloshinskii, I. E. Sov. Phys. JETP 5, 1259 (1957).

34. Heinze, S. et al. Spontaneous atomic-scale magnetic skyrmion lattice in two dimensions. Nature Phys. 7, 713–718 (2011).

35. Romming, N. et al. Writing and deleting single magnetic skyrmions. Science 341, 636–639 (2013).

36. Romming, N., Kubetzka, A., Hanneken, C., von Bergmann, K. & Wiesendanger, R. Field-dependent size and shape of single magnetic skyrmions. Phys. Rev. Lett. 114, 177203 (2015).

37. Fert, A. Magnetic and transport properties of metallic multilayers. Mater. Sci. Forum 59–60, 439–480 (1990).

38. Soumyanarayanan, A. et al. Tunable room temperature magnetic skyrmions in Ir/Fe/Co/Pt multilayers. Nature Mater. 16, 898-904 (2017).

39. Pollard, S. D. et al. Observation of stable Néel skyrmions in cobalt/palladium multilayers with Lorentz transmission electron microscopy. Nature Commun. 8, 14761 (2017).

40. Schott, M. et al. The skyrmion switch: turning magnetic skyrmion bubbles on and off with an electric switch. Nano Lett. 17, 3006–3012 (2017).

41. Yu, G. et al. Room-temperature creation and spin-orbit torque manipulation of skyrmions in thin films with engineered asymmetry. Nano Lett. 16, 1981–1988 (2016).

42. Jiang, W. et al. Direct observation of the skyrmion Hall effect. Nature Phys. 13, 162–169 (2016).

43. Hrabec, A. et al. Current-induced skyrmion generation and dynamics in symmetric bilayers. Nature Commun. 8, 15765 (2017).

44. Litzius, K. et al. Skyrmion Hall effect revealed by direct time-resolved X‑ray microscopy. Nature Phys. 13, 170–175 (2017).

45. Yu, G. Q. et al. Room-temperature skyrmion shift device for memory application. Nano Letters 17, 261–268 (2017).

46. Legrand, W. et al. Room-temperature current-induced generation and motion of sub‑100 nm skyrmions. Nano Lett. 17, 2703–2712 (2017).

47. Tomasello, R. et al. A strategy for the design of skyrmion racetrack memories. Sci. Rep. 4, 6784 (2014).

48. Huang, Y. et al. Magnetic skyrmion-based synaptic devices. Nanotechnology 28, 08LT02 (2017).

49. Prychynenko, D. et al. Magnetic skyrmion as a nonlinear resistive element: a potential building block for reservoir computing. Phys. Rev. Applied 9, 014034 (2018).

50. Jude, R. et al. Magnetic skyrmions in confined geometries: Effect of the magnetic field and the disorder. J. Magn. Magn. Mater. 455, 3-8, (2018).

51. Büttner, F. et al. Field-free deterministic ultrafast creation of magnetic skyrmions by spin–orbit torques. Nature Nanotech. 12,1040–1044 (2017).

52. Woo, S. et al. Deterministic creation and deletion of a single magnetic skyrmion observed by direct time-resolved X-ray microscopy. Nature Electron. 1, 288-296 (2018).

53. Mangin, S. et al. Engineered materials for all-optical helicity-dependent magnetic switching. Nature Mater. 13, 286-292 (2014).

54. Lambert, C-H. et al. All-optical control of ferromagnetic thin films and nanostructures. Science 345, 1337 (2014).

55. Vallobra, P. et al. Manipulating exchange bias using all-optical helicity-dependent switching. Phys. Rev. B 96, 144403 (2017).

56. El Hadri M. S. et al. Two types of all-optical magnetization switching mechanisms using femtosecond laser pulses. Phys. Rev. B 94, 064412 (2016).

57. El Hadri M. S. et al. Domain size criterion for the observation of all-optical helicity-dependent switching in magnetic thin films. Phys. Rev. B 94, 064419 (2016).

58. Quessab, Y. et al. Helicity-dependent all-optical domain wall motion in ferromagnetic thin films. Phys. Rev. B 97, 054419 (2018).

59. Koshibae, W. & Nagaosa N. Creation of skyrmions and antiskyrmions by local heating. Nature Commun. 5, 5148 (2014).

60. Finazzi, M. et al. Laser-induced magnetic nanostructures with tunable topological properties. Phys. Rev. Lett. 110, 177205 (2013).

61. Berruto, G. et al. Laser-Induced Skyrmion Writing and Erasing in an Ultrafast Cryo-Lorentz Transmission Electron Microscope. Phys. Rev. Lett. 120, 117201 (2018).

62. Srivastava, T. et al. Large voltage tuning of Dzyaloshinskii-Moriya Interaction: a route towards dynamic control of skyrmion chirality. Nano Lett. 18, 4871-4877 (2018)..

63. Saratz, N., Ramsperger, U., Vindigni, A. & Pescia, D. Irreversibility, reversibility, and thermal equilibrium in domain patterns of Fe films with perpendicular magnetization. Phys. Rev. B 82, 184416 (2010).

64. Fujita, H. & Sato, M. Ultrafast generation of skyrmionic defects with vortex beams: Printing laser profiles on magnets. Phys. Rev. B 95, 054421 (2017).

65. Lin, S. Z., Reichhardt, C., Batista, C. D. & Saxena, A. Particle model for skyrmions in metallic chiral magnets: dynamics, pinning, and creep. Phys. Rev. B 87, 214419 (2013).

66. Zhang, X. et al. Skyrmion–skyrmion and skyrmion–edge repulsions in skyrmion-based racetrack memory. Sci. Rep. 5, 7643 (2015).

67. Thiele A. A. The theory of cylindrical magnetic domains, Bell Syst. Tech. J. 48, 3287 (1969)

68. Hubert, A. & Schäfer, R. Magnetic Domains: The Analysis of Magnetic Microstructures (Springer, 1998).

69. Donahue, M. J. & Porter, D. G. OOMMF User's guide, Version 1.0, Interagency Report NIST IR 6376 (Gaithersburg, MD (1999).

70. Rohart, S. & Thiaville, A. Skyrmion confinement in ultrathin film nanostructures in the presence of Dzyaloshinskii–Moriya interaction. Phys. Rev. B 88, 184422 (2013).

71. Koopmans, B. et al. Explaining the paradoxical diversity of ultrafast laser-induced demagnetization. Nature Mater. 9, 259–265 (2010).

72. Cornelissen, T. D., Córdoba, R. & Koopmans, B. Microscopic model for all optical switching in ferromagnets, Appl. Phys. Lett. 108, 142405 (2016).

73. Brächer, T. et al. Detection of short-waved spin waves in individual microscopic spin-wave waveguides using the inverse spin Hall effect. Nano Lett. 17, 7234-7241 (2017).

74. Yin, G. et al. Topological charge analysis of ultrafast single skyrmion creation. Phys. Rev. B 93, 174403 (2016).

75. Benitez, M. J. et al. Magnetic microscopy and topological stability of homochiral Néel domain walls in a Pt/Co/AlO$_x$ trilayer. Nature Commun. 6, 8957 (2015).

76. Rohart, S., Miltat, J. & Thiaville, A. Path to collapse for an isolated Néel skyrmion. Phys. Rev. B 93, 214412 (2016).

77. F. Büttner, I. Lemesh & G. S. D. Beach, Theory of isolated magnetic skyrmions: From fundamentals to room temperature applications, Sci. Rep. 8, 4464 (2018).

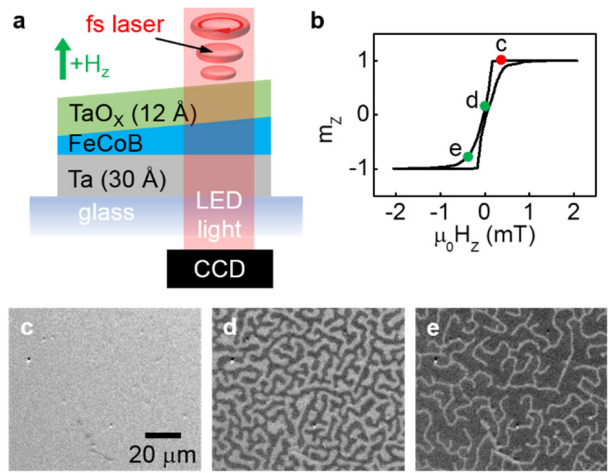

**Figure 1. Ultrathin ferromagnetic film structure and domain patterns. a.** Schematic representation of the film structure and the experimental setup. The film is excited by fs laser pulses, and a 680-nm monochromatic LED (light-emitting diode) light is used to image magnetic domains. **b.** Evolution of the normalized perpendicular component of the magnetization $m_z$ with respect to the perpendicular field $H_z$. **c-e** Magnetic domain patterns obtained at the points c, d and e as marked in Fig. 1b.

**Fig. 1** *Je et al.*

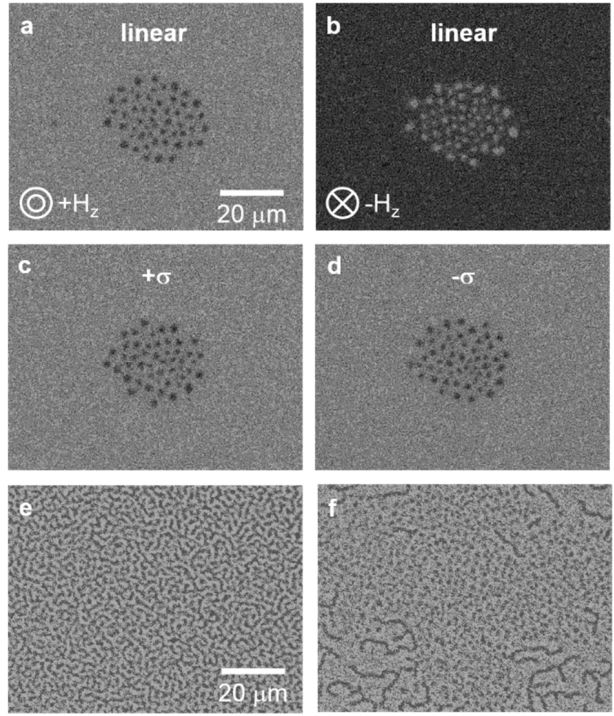

**Figure 2. Femto-second laser-induced creation of skyrmion bubbles. a.** Nucleated bubbles from a saturated state (+z) by a 35-fs single laser pulse with a laser fluence of 1.84 mJ/cm$^2$ and linear polarization. **b.** Bubble creation from an oppositely saturated state (-z) with linearly polarized light. **c-d.** Bubble creation by a single laser with right-handed $\sigma^+$ polarization (c) and left-handed $\sigma^-$ polarization (d). The dc magnetic field is +0.31 mT for a, c and d and -0.31 mT for b. The magnetization of the bubble centre is antiparallel to the field. **e-f.** Much narrower stripe domains for a thicker FeCoB (~10 Å) at zero field (e) and created bubbles smaller than 1 μm in diameter under the $\mu_0 H_z$ of +0.7 mT (f). Particularly for f, the laser was swept over the film to increase the number of bubbles for visibility.

**Fig. 2** *Je et al.*

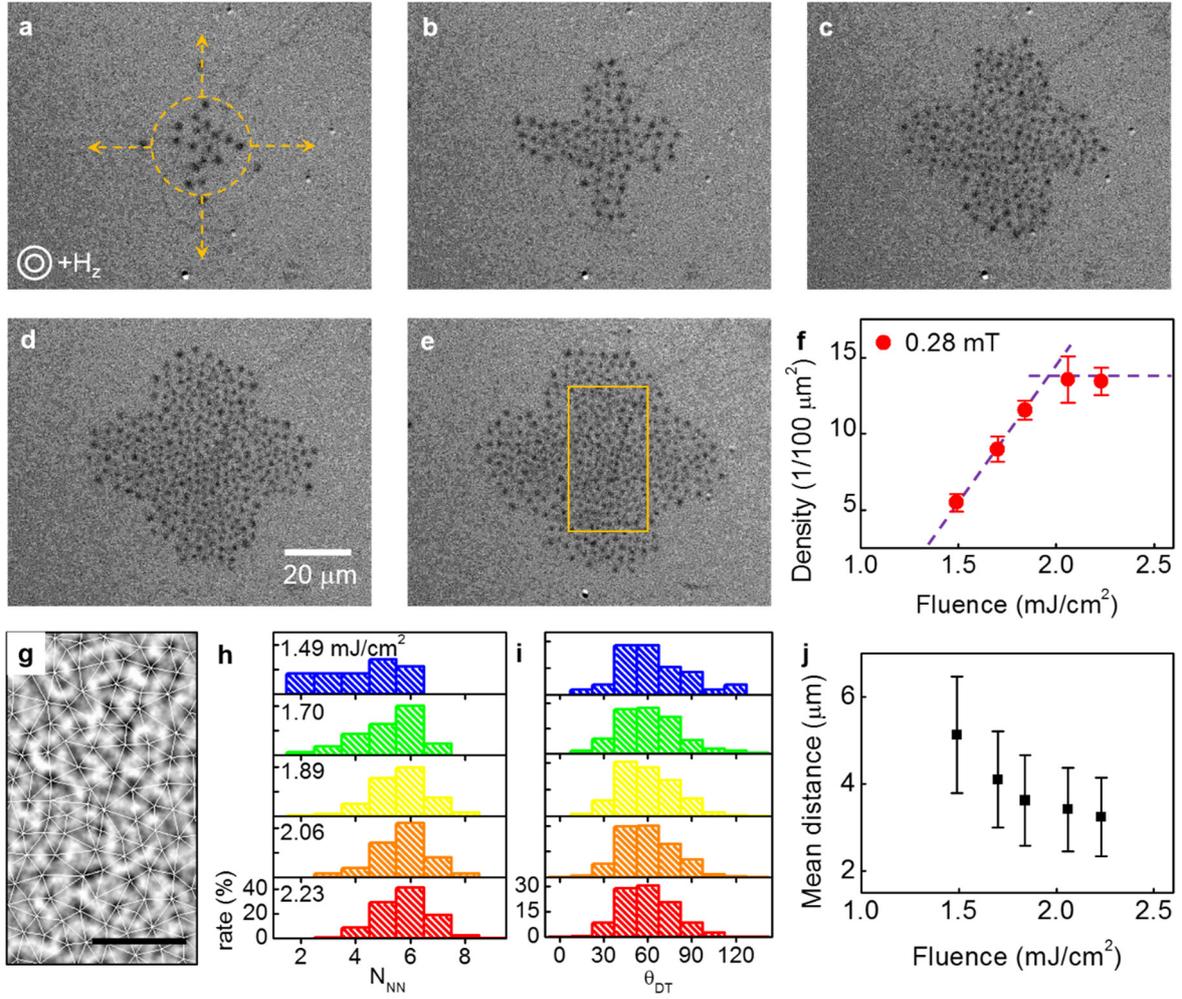

**Figure 3. Laser fluence dependence of the bubble and bubble lattice creation. a-e.** Magnetic images of ensemble of bubbles made by sweeping laser with laser fluences of 1.49 mJ/cm$^2$ (a), 1.70 mJ/cm$^2$ (b), 1.84 mJ/cm$^2$ (c), 2.06 mJ/cm$^2$ (d) and 2.23 mJ/cm$^2$ (e) in the presence of $\mu_0 H_z$ = +0.28 mT. The orange rectangular marks the same area with (g). **f.** Bubble density with respect to the laser fluence. We covered the bubble area with square boxes with lateral sizes between 10μm and 20μm and we only counted the bubbles present in the squares. The dashed lines are guides for the eye. **g.** A typical example of Delaunay triangularization cropped from the rectangular in Fig. 3e. The scale bar is 10 μm. A band-pass filtering and mean filtering are used to enhance the magnetic contrast and thus to capture the centres of bubbles using a 2D Gaussian fit. **h-i.** The statistics of the nearest neighbour number $N_{NN}$ and the angle of Delaunay triangles $\theta_{DT}$ obtained from Delaunay triangularization. Laser fluence increases from the top panel to the bottom panel. **j.** The mean distance between nearest bubbles as a function of laser fluence.

**Fig. 3** *Je et al.*

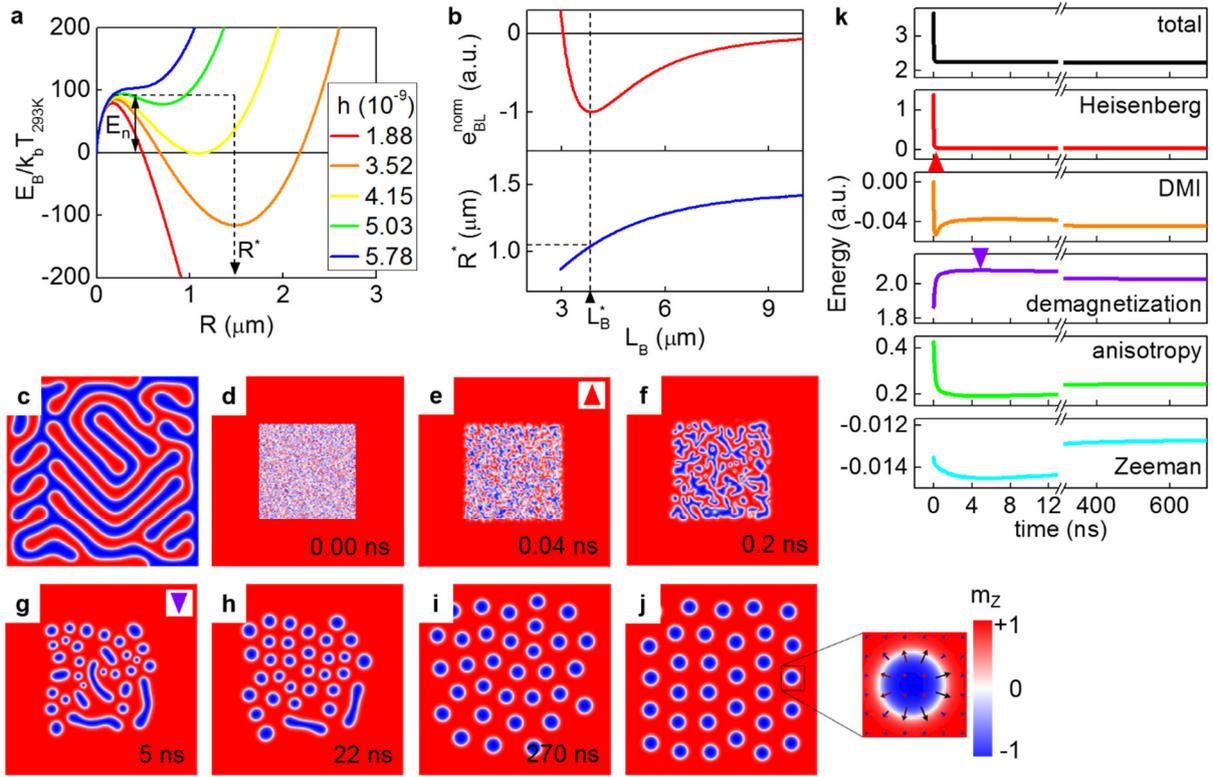

**Figure 4. Analytic models and micromagnetic simulations for bubble and bubble lattice creation. a.** Free energy of the isolated bubble as a function of the bubble radius R. $E_n$ stands for the energy barrier for nucleation of the micro-size bubble. **b.** Normalized free energy of a hexagonal bubble lattice as a function of the bubble spacing $L_B$ (upper panel) and the radius of bubbles (lower panel) which minimizes the bubble lattice energy for a given $L_B$. The calculation is performed for $\mu_0 H_z = +0.28$ mT. Dashed vertical line represents the minimum energy point of the bubble lattice, resulting in the saturation of the bubble density in Fig. 3f. **c.** A labyrinthine domain pattern obtained using a micromagnetic simulation at zero field. **d-j.** The generation process of the skyrmion lattice. The simulation area is 2 μm × 2 μm and a randomly oriented magnetization state is embedded in the central area of 1 μm × 1 μm as an initial magnetization state (d). The simulation is performed in the presence of a magnetic field of +5 mT. The images show domain morphology with respect to the simulation time. Inset of (j) details a single skyrmion structure and the lateral size of the image is 240 nm. **k.** Time evolution of each energy term. The red triangle (purple triangle) corresponds to e (g).

**Fig. 4** *Je et al.*

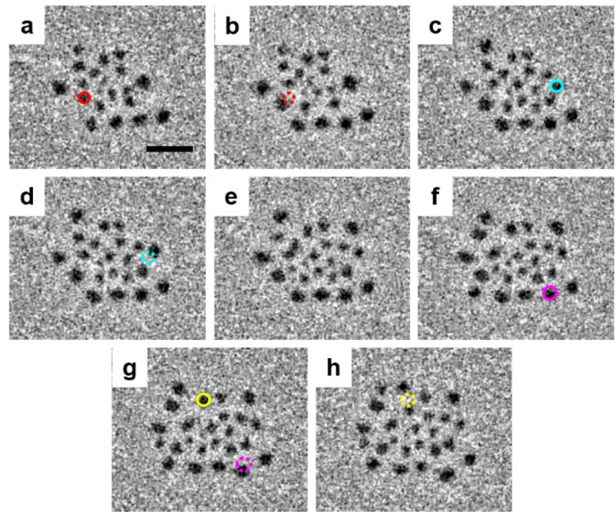

**Figure 5. Repulsive bubble-bubble interaction. a-h.** Sequence of images taken after every single laser pulse. The circles with a dashed line indicate their previous positions labelled as circles with a solid line. The applied field $\mu_0 H_z$ is +0.28 mT. The scale bar is 10 μm.

**Fig. 5** *Je et al.*

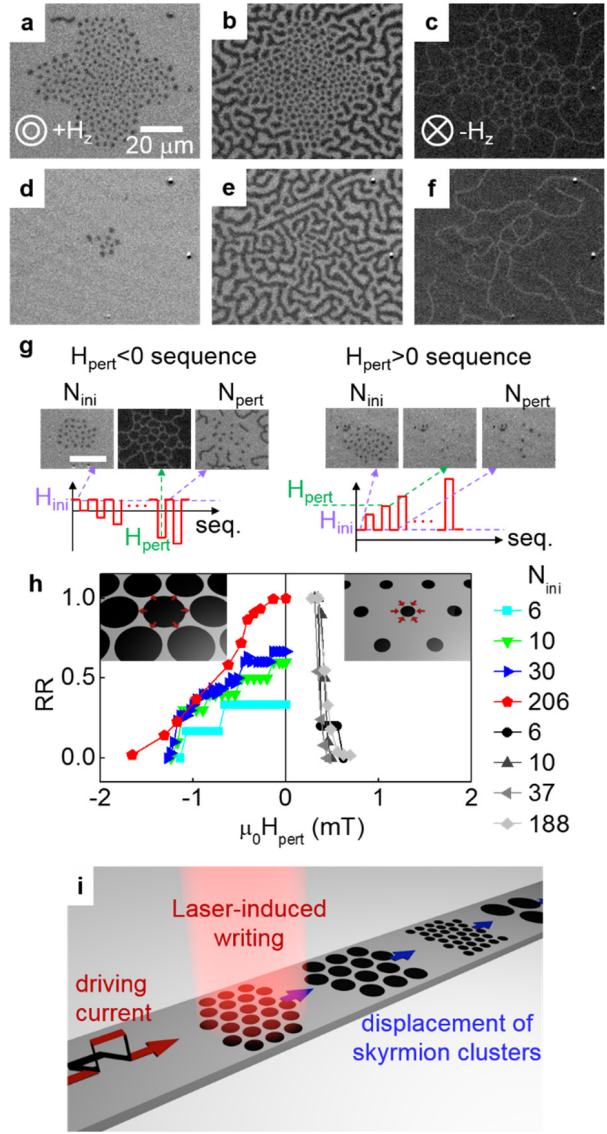

**Figure 6. Response of ensembles of bubbles to the external magnetic field variation. a-c.** Change of magnetic domain morphology of a bubble lattice with respect to $\mu_0 H_z$. Initially, the bubble lattice is generated by laser pulses in the presence of $\mu_0 H_z$ of +0.28 mT (a). Applied magnetic fields for b and c are 0 and -1 mT, respectively. **d-f.** Change of magnetic domain morphology of sparsely distributed bubbles with respect to $\mu_0 H_z$. The magnetic fields for d, e and f are +0.28, 0 and -1 mT, respectively. **g.** Schematic representation of the sequence of the magnetic perturbation for the negative $H_{pert}$ and positive $H_{pert}$. The scale bar is 20 µm. **h.** Bubble retention ratio $RR$ with respect to $H_{pert}$. The legend on the right shows $N_{ini}$. The insets illustrate the experimental situations for the negative (expansion) and positive (shrinking) $H_{pert}$ sequences. **i.** Schematic of the skyrmion lattice shift register where information is encoded in the form of ensembles of skyrmions of different skyrmion densities. The density is controlled by laser fluence. Once a skyrmion cluster is written by laser, clusters including newly written one can be shifted (blue arrows) by the current pulse flowing through the track, resulting the motion of trains of skyrmion clusters.

**Fig. 6** *Je et al.*